\documentclass[prl,letterpaper,twocolumn,showpacs,superscriptaddress]{revtex4}
\usepackage{graphicx,psfrag,amsmath,amssymb,amsfonts,bbm,latexsym,color,dcolumn}
\begin{document}

\title{Cooling dynamics of ultracold two-species Fermi-Bose mixtures}

\author{Carlo Presilla}
\affiliation{Dipartimento di Fisica, Universit\`a di Roma ``La Sapienza'',
Piazzale A. Moro 2, Roma 00185, Italy}
\affiliation{Istituto Nazionale per la Fisica della Materia, 
Unit\`a di Roma 1 and Center for Statistical Mechanics and Complexity,
Roma 00185, Italy}
\affiliation{Istituto Nazionale di Fisica Nucleare, Sezione di Roma 1, 
Roma 00185, Italy}
\author{Roberto Onofrio}
\affiliation{Dipartimento di Fisica ``G. Galilei'',
 Universit\`a di Padova, Via Marzolo 8, Padova 35131, Italy}
\affiliation{Istituto Nazionale per la Fisica della Materia, 
Unit\`a di Roma 1 and Center for Statistical Mechanics and Complexity, 
Roma 00185, Italy}
\affiliation{Los Alamos National Laboratory, Los Alamos, New Mexico 87545}
\date{\today}

\begin{abstract}
We compare strategies for evaporative and sympathetic cooling of
two-species Fermi-Bose mixtures in single-color and two-color 
optical dipole traps. 
We show that in the latter case a large heat capacity of the bosonic
species can be maintained during the entire cooling process. 
This could allow to efficiently achieve a deep Fermi degeneracy regime having 
at the same time a significant thermal fraction for the Bose 
gas, crucial for a precise thermometry of the mixture. 
Two possible signatures of a superfluid phase transition for 
the Fermi species are discussed.
\end{abstract}

\maketitle
Recent studies of ultracold dilute matter are bridging a gap between
the idealized descriptions of quantum degenerate Bose and Fermi gases
and their actual counterpart in strongly-interacting condensed matter 
systems like liquid $^4$He and electrons in superconducting materials 
\cite{Pethick}. 
While experimental studies of interacting dilute Bose gases in the
degenerate regime are ongoing since 1995  \cite{Wieman},  
Fermi gases have been explored only more recently. 
Non-interacting, purely quantum-mechanical features of dilute Fermi
gases have been observed in the degenerate regime, namely Pauli
blocking \cite{DeMarco} and Fermi pressure \cite{Truscott,Schreck}.  
Phenomena involving their interacting nature are expected when fermions 
are highly degenerate. Important studies of strongly interacting degenerate
Fermi gases have been recently reported for Fermi-Bose mixtures
\cite{Inguscio}, and for two-component Fermi gases \cite{Oharanew}. 
Moreover, BCS-based models are predicting 
a superfluid phase based on Cooper pairing already invoked for the 
understanding of low-temperature superconductivity and superfluidity
of $^3$He \cite{Stoof}. 

Current efforts to cool fermions seem 
ultimately limited by intrinsic heating sources \cite{Timmermans} in
the case of evaporative cooling of two hyperfine states of fermions 
\cite{DeMarco,Granade}, and by the decreasing cooling efficiency of bosons 
in experiments using sympathetic cooling  
\cite{Truscott,Schreck,Hadzibabic,Roati}. 
Recently, we proposed the use of a two-color optical dipole trap 
to enhance the Fermi degeneracy temperature $T_\mathrm{F}$ with
respect to the Bose-Einstein critical temperature $T_\mathrm{c}$
whenever a bosonic species is used to sympathetically cool a different 
fermionic species \cite{Onofrio}. 
This can be obtained by engineering different trapping potentials for 
the two species with proper detuning and intensities of two laser
beams, as discussed in \cite{Onofrio} with particular regard 
to the static confinement features. In this Letter we analyze the dynamics
of cooling and heating of the Fermi-Bose mixture in two-color optical dipole
traps. It turns out that a deep Fermi degeneracy regime can be
achieved, allowing at the same time for both precision thermometry 
and relatively simple signatures of fermion superfluidity.

In order to understand the efficiency limits of sympathetic cooling 
of a fermion-boson mixture let us consider the heat capacities $C(N,T)$
of the two species at a fixed number $N$ of particles
as a function of temperature $T$.
In the case of non-interacting gases confined into a harmonic 
potential $V(x,y,z) = \frac{1}{2} m \left( \omega_x^2 x^2 + \omega_y^2
y^2 + \omega_z^2 z^2 \right)$, the heat capacities can be evaluated numerically. 
In Fig. 1 we show the behavior of $C_\mathrm{b}(N_\mathrm{b},T)$ and 
$C_\mathrm{f}(N_\mathrm{f},T)$ for a mixture composed by 
the same number of bosons and fermions trapped in a 
crossed-beam optical dipole trap \cite{Adams,Chapman}.
Below the critical, 
$T_\mathrm{c} = \zeta(3)^{-1/3}\hbar \omega_\mathrm{b} 
N_\mathrm{b}^{1/3} k_\mathrm{B}^{-1}$,
and Fermi, $T_\mathrm{F} = 6^{1/3}\hbar \omega_\mathrm{f}
N_\mathrm{f}^{1/3} k_\mathrm{B}^{-1}$,
temperatures, defined in terms of the average angular trap frequencies 
$\omega_\mathrm{b}=
(\omega_{\mathrm{b}x} \omega_{\mathrm{b}y} \omega_{\mathrm{b}z})^{1/3}$
and  
$\omega_\mathrm{f}=
(\omega_{\mathrm{f}x} \omega_{\mathrm{f}y} \omega_{\mathrm{f}z})^{1/3}$,
the boson and fermion heat capacities vanish as $T^3$ and $T$,
respectively.
If $\omega_\mathrm{f}=\omega_\mathrm{b}$, 
the boson heat capacity becomes smaller than the fermion 
one below $T/T_\mathrm{F} \simeq 0.3$, strongly affecting the  
efficiency of sympathetic cooling for smaller $T/T_\mathrm{F}$.
This explains qualitatively the difficulty in reaching temperatures 
lower than $T/T_\mathrm{F}\simeq 0.25$ in the experiments
reported in \cite{Truscott,Schreck} where $^7$Li-$^6$Li mixtures 
were used, and more in general in magnetically or single-color optically 
trapped Fermi-Bose mixtures. 
\begin{figure}[t]
\vspace{-5mm}
\psfrag{w_f/w_b=10}{$\omega_\mathrm{f}/\omega_\mathrm{b}=10$}
\psfrag{w_f/w_b=1}{$\omega_\mathrm{f}/\omega_\mathrm{b}=1$}
\psfrag{T/T_F}{$T/T_\mathrm{F}$}
\psfrag{C_N/Nk_B}{$C/Nk_\mathrm{B}$}
\includegraphics[width=0.95\columnwidth]{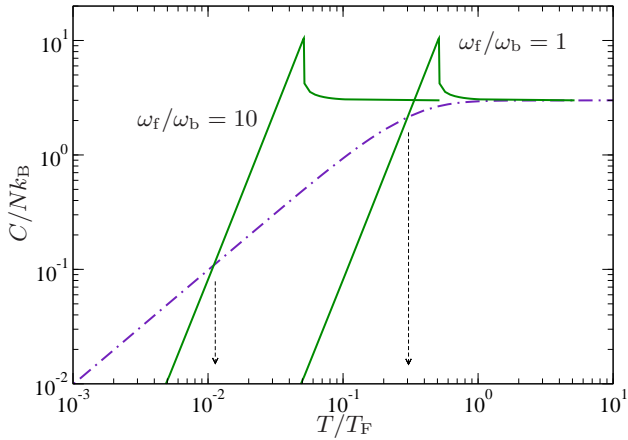}
\caption{
Heat exchange for harmonically trapped Bose-Fermi mixtures.
The single particle heat capacity of non-interacting fermions (dot-dashed)
and bosons (solid) is shown versus temperature for two different values of the 
trap frequency ratio $\omega_\mathrm{f}/\omega_\mathrm{b}$.
Arrows evidence the different $T/T_F$ values below which 
the boson heat capacity becomes smaller than the fermion one.
We consider a case with $N_\mathrm{b}=N_\mathrm{f}=10^6$, $m_\mathrm{b}=m_\mathrm{f}$,
and $\omega_x=\omega_y=\omega_z/\sqrt{2}$.
}
\label{fig1}
\end{figure}
Were we able to increase the ratio $\omega_\mathrm{f}/\omega_\mathrm{b}$, 
the cooling efficiency of a boson-fermion mixture could be extended to 
much lower temperatures.  As an example, in Fig. 1 we show that for 
$\omega_\mathrm{f}/\omega_\mathrm{b}=10$, obtainable with bichromatic 
optical dipole traps \cite{Onofrio}, the heat capacity inversion
takes place at $T/T_\mathrm{F}\simeq 10^{-2}$. 
The discussion can be made more quantitative by considering the dynamics of 
the system during forced evaporation and comparing a bichromatic trap
to the single-color case. 

In optical dipole traps forced evaporative cooling is obtained by
continuously decreasing the depth of the confining potential energy 
via proper control 
of the laser power. The detailed dynamics for a single-color optical 
dipole trap has been discussed in \cite{OHara}, where scaling laws 
for all the relevant parameters of evaporative cooling were obtained. 
A fundamental quantity in forced evaporation is the ratio between 
the potential energy depth experienced by the trapped atoms and their 
temperature, $\Delta U/k_\mathrm{B}T \equiv \eta$. 
It has been shown that thermodynamic equilibrium is assured if $\eta$ 
is kept constant, even if $\Delta U$ and $T$ are time dependent 
\cite{Ketterle}. This implies that during cooling the temperature 
of the atomic cloud is always well-defined. 
The condition of constant $\eta$  determines the time dependence of the 
potential energy depth \cite{OHara}
\begin{equation}
\label{scaling}
\frac{\Delta U(t)}{\Delta U_i} = 
\left(1+\frac{t}{\tau}\right)^{\varepsilon_U},
\end{equation}
where $\Delta U_i$ is the initial potential depth,  
$\varepsilon_U=-2(\eta^\prime-3)/\eta^\prime$,
and $\tau^{-1}=(2/3) \eta^\prime(\eta-4)\exp(-\eta)\gamma_i$, 
with $\eta^\prime=\eta+(\eta-5)/(\eta-4)$ and $\gamma_i$ 
being the initial elastic collision rate. 
Once the time dependence of $\Delta U$ is known, all other relevant
quantities, {\it e.g.} number of particles, temperature, phase space density, 
and elastic scattering rate, are obtained by scaling laws 
similar to (\ref{scaling}) with possibly different exponents,
$\varepsilon_N$, $\varepsilon_T$, $\varepsilon_\rho$, and $\varepsilon_\gamma$.
In Table I we report the values of these exponents for three different 
values of $\eta$ realistically achievable in experimental situations, 
and the corresponding time constant $\tau$.
\begin{table}
\vspace{-5mm}
\begin{ruledtabular}
\begin{tabular}{cccccc}
$\eta$ & $\varepsilon_U$, $\varepsilon_T$ & $\varepsilon_N$ &
  $\varepsilon_\rho$ & 
$\varepsilon_\gamma$ & $(\gamma_i \tau)^{-1}$ \\ \hline 
 5 & $-0.80$ & $-0.60$ & 0.60 & $-1$ & $2.2 \times 10^{-2}$ \\ 
10 & $-1.45$ & $-0.28$ & 1.89 & $-1$ & $2.0 \times 10^{-3}$ \\ 
15 & $-1.62$ & $-0.19$ & 2.25 & $-1$ & $3.6 \times 10^{-5}$ \\
\end{tabular}
\end{ruledtabular}
\caption{Evaporative cooling scaling exponents for the potential energy depth 
$\varepsilon_U$, the temperature $\varepsilon_T$, 
the number of particles $\varepsilon_N$, 
the phase-space density $\varepsilon_\rho$, 
and the elastic collision rate $\varepsilon_\gamma$, 
for three values of the evaporation parameter $\eta$. 
We also report the time constant 
$\tau$ in terms of the initial elastic scattering rate 
$\gamma_i=N_\mathrm{b}m_\mathrm{b} \sigma \omega_\mathrm{b}^3/(2\pi^2
k_\mathrm{B} T)$, where 
$N_\mathrm{b}$, $\omega_\mathrm{b}$ and $T$ are the initial values of
the number of bosons, their average angular 
trap frequency and temperature, respectively,  
and $\sigma$ the elastic cross-section.}
\end{table}

In the case of a mixture of bosonic and fermionic species,
Eq. (\ref{scaling}) describes the potential energy depth of bosons
$\Delta U_\mathrm{b}$ \cite{NOTE}.
This quantity, in turns, fixes the laser power $P$ 
necessary to create the confining potential well. 
The potential energy depth of fermions $\Delta U_\mathrm{f}$ is
then determined as a function of $P$. By using the scaling law 
(\ref{scaling}), we have studied evaporative cooling strategies 
for single-color and two-color optical dipole traps with 
$^6$Li-$^{23}$Na mixtures, recently brought to quantum degeneracy for 
both species in a magnetic trap \cite{Hadzibabic}. 
Analogous  considerations hold for all the combinations of the 
only two available stable Fermi alkali isotopes, $^6$Li and $^{40}$K, 
refrigerated through the largely available Bose coolers, $^{23}$Na and $^{87}$Rb. 
In the single-color case only a red-detuned laser beam is present
and its power $P_1$ is decreased continuously as shown by the 
dashed line in the upper panel of Fig. 2. 
The ratio of the fermionic and bosonic average trapping frequencies
has the constant value $\omega_\mathrm{f}/\omega_\mathrm{b} = 1.96$ 
determined by the masses of the two species and the different detunings of the atomic 
transition wavelengths with respect to the red-detuned laser wavelength.
In the two-color situation a coaxial blue-detuned beam focused on the center 
of the existing optical dipole trap is turned on at time $t_0$
and for $t \geq t_0$ its power $P_2$ is maintained at a constant ratio
with the red-detuned laser power, $P_2/P_1 = \mathrm{constant}$. 
By demanding a smooth time-dependence for $\Delta U_\mathrm{b}(t)$ 
as described by Eq. (\ref{scaling}) -
see central panel of Fig. 2 - together with the fact that 
$\Delta U_\mathrm{b} = \Delta U_\mathrm{b} (P_1,P_2)$,
a discontinuity at $t=t_0$ also for the red-detuned laser power is required.
For $t\geq t_0$ the ratio $\omega_\mathrm{f}/\omega_\mathrm{b}$ can be
ideally increased to an arbitrary high value by choosing a proper ratio
$P_2/P_1$.
\begin{figure}[t]
\vspace{-5mm}
\psfrag{P_1}{$P_1$}
\psfrag{P_1}{$P_1$}
\psfrag{P_2}{$P_2$}
\psfrag{powers}{$P$ (W)}
\psfrag{DU_f}{$\Delta U_\mathrm{f}$}
\psfrag{DU_f}{$\Delta U_\mathrm{f}$}
\psfrag{DU_b}{$\Delta U_\mathrm{b}$}
\psfrag{trap depths}{$\!\!\!\! \Delta U/k_\mathrm{B}$ ($\mu K$)}
\psfrag{w_f}{$\omega_\mathrm{f}$}
\psfrag{w_f}{$\omega_\mathrm{f}$}
\psfrag{w_b}{$\omega_\mathrm{b}$}
\psfrag{w_b}{$\omega_\mathrm{b}$}
\psfrag{time}{$\gamma_i t$}
\psfrag{trap frequencies}{$\omega/2\pi$ (kHz)}
\includegraphics[width=0.95\columnwidth]{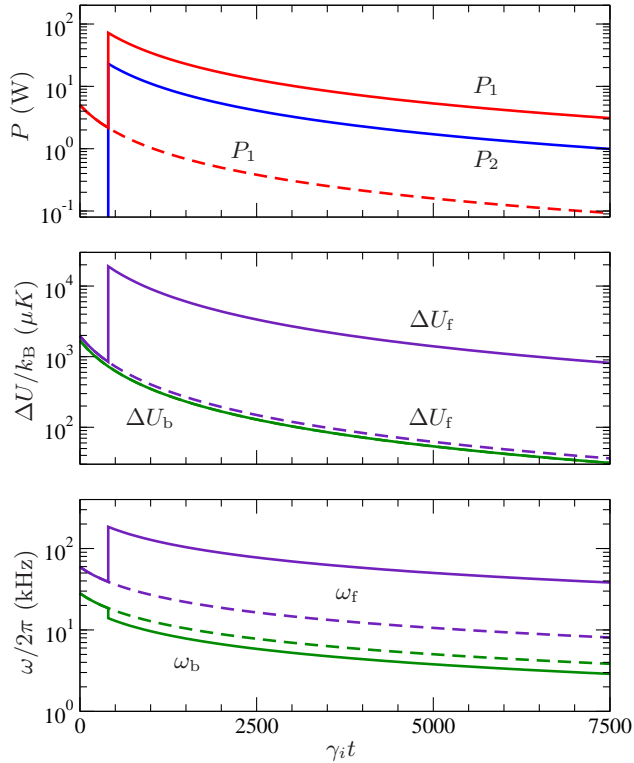}
\caption{
Evaporative cooling strategies for an optically trapped 
$^6$Li-$^{23}$Na mixture. Time evolution of red-detuned 
and blue-detuned laser powers (upper panel), fermion and boson 
trap depths (central panel) and fermion and boson trap frequencies (lower panel)
for a single-color optical dipole trap (dashed lines) 
and a two-color optical dipole trap (solid lines).
The laser powers are fixed by the condition that  
$\Delta U_\mathrm{b}(t)$ follows Eq. (\ref{scaling}) with $\eta=10$
and $P_2/P_1=0$ in the single-color case or $P_2/P_1=0.32$ 
(corresponding to $\omega_\mathrm{f}/\omega_\mathrm{b} = 13.15$) 
for $\gamma_i t \geq 400$ in the two-color case. 
The wavelengths of the laser beams are chosen at $\lambda_1=$1064 nm and $\lambda_2$=532 nm.
}
\label{fig2}
\end{figure}
The abrupt increase (decrease) of the fermionic (bosonic) average
trapping frequency due to the turning-on of the blue-detuned laser
determines a corresponding decrease (increase) of the Fermi (critical) 
temperature. As shown in the lower panel of Fig. 3, this produces an 
increase of $T/T_\mathrm{c}$ and a decrease of $T/T_\mathrm{F}$ 
with respect to their corresponding smooth evolutions in the case 
of a single-color trap. 
Therefore, the presence of the blue-detuned laser helps to both maintain
the Bose gas in a non-condensed state and allow for a deeper
degeneracy condition of the Fermi gas.
At the same time, evaporative cooling is less efficient as a
consequence of the weakening of the confinement caused by the 
blue-detuned beam.
However, this is not an issue since in the latest stage of evaporation 
we estimate elastic scattering rates of order 
$\Gamma_\mathrm{el}\simeq 10^2~\mathrm{s}^{-1}$ even taking into
account the suppression of scattering induced by Pauli blocking \cite{Holland0}. 
Also, the blue-detuned beam gives the dominant contribution to the 
heating of the mixture due to Rayleigh scattering of the sodium atoms,
but this is largely compensated by the cooling power of the Bose gas
even in the latest stage of the evaporation.
Other sources of heating, like hole heating \cite{Timmermans} or
technical laser noise \cite{Gehm}, can be made negligible with respect to the heating
induced by Rayleigh scattering. In particular, fluctuations in the laser power ratio
$P_2/P_1$ could induce instabilities and parametric heating  
especially in the interesting regime where the ratio
$\omega_\mathrm{f}/\omega_\mathrm{b}$ is made large. 
For $\omega_\mathrm{f}/\omega_\mathrm{b}=10$, a power ratio stability 
of 0.1-1$\%$ is required, which is within the capability of current
laser stabilization techniques. 

In the upper panel of Fig. 3 we show the evolution of the ratio 
$C_\mathrm{b}/C_\mathrm{f}$ between the heat capacities of the bosonic
and fermionic species - a figure of merit of the sympathetic cooling efficiency.
The heat capacities have been evaluated numerically taking 
into account the 
time evolution of the relevant quantities, in particular
the diminishing number of bosons during forced evaporation which 
strongly affects the time-dependence of $C_\mathrm{b}$
(while the effect of many-body interactions, 
evaluated in \cite{Giorgini}, gives a much weaker time-dependence). 
The boson heat capacity in the single-color trap becomes smaller 
than the fermion one at times $\gamma_i t \gtrsim 6200$ when 
most of the bosons are condensed ($T/T_\mathrm{c} \simeq 0.25$, 
$T/T_\mathrm{F} \simeq 0.1$).
On the other hand, in the two-color trap $C_\mathrm{b}/C_\mathrm{f}$
maintains values much larger than unity and it is possible to reach 
$T/T_\mathrm{F} \simeq 0.02$ while $T/T_\mathrm{c} \gtrsim 0.3$. 
\begin{figure}[t]
\vspace{-5mm}
\psfrag{T/T_F}{$T/T_\mathrm{F}$}
\psfrag{T/T_c}{$T/T_\mathrm{c}$}
\psfrag{C_b/C_f}{$C_\mathrm{b}/C_\mathrm{f}$}
\psfrag{T/T_F T/T_c}{$T/T_\mathrm{F}$, $T/T_\mathrm{c}$}
\psfrag{time}{$\gamma_i t$}
\includegraphics[width=0.95\columnwidth]{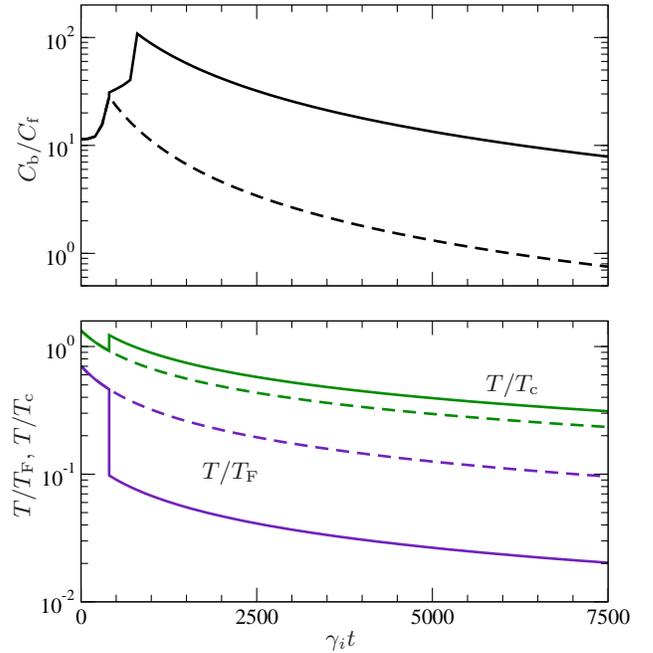}
\caption{Efficiency of sympathetic cooling in optical dipole traps.
The time evolution of heat capacity ratio $C_\mathrm{b}/C_\mathrm{f}$
(upper panel) and temperature ratios
$T/T_\mathrm{F}$ and $T/T_\mathrm{c}$ (lower panel) are depicted 
for a $^6$Li-$^{23}$Na mixture in which initially 
$N_\mathrm{f}=10^5$ and $N_\mathrm{b}=10^6$.
The single-color (dashed lines) and two-color (solid lines)
refer to the trapping configurations  defined in Fig. 2. 
Unlike the latter, for the single-color case the 
heat capacity ratio approaches unity in the latest stage, and 
the equilibrium temperature of the mixture is no longer dominated 
by the bosonic component undergoing forced evaporation.
}
\label{fig3}
\end{figure}
The presence of a larger bosonic thermal cloud for the two-color case 
even at the latest stage of fermion cooling allows for a more precise thermometry.  
The estimate of the temperature for the Fermi-Bose mixture is indeed 
obtained by fitting the tail of the normal Bose component 
superimposed to the condensate fraction. 
As the temperature is lowered the thermal component shrinks in amplitude and size  
therefore lowering the accuracy of the measurement.
This effect is mitigated in the two-color trap. 

Superfluidity of the Fermi gas is expected below the
critical temperature for the onset of atomic Cooper pairs \cite{Stoof} 
$T_\mathrm{BCS} \simeq 5/3  \exp(-\pi/2k_\mathrm{F}|a|-1) \,
T_\mathrm{F}$, where $k_\mathrm{F}$ is
the Fermi wavevector and $a$ the elastic scattering length of fermions. 
Besides leaving freedom to apply arbitrary homogeneous magnetic fields
to enhance the scattering length through tuning to a Feshbach
resonance \cite{Timmermans2001,Holland,Ohashi}, 
our bichromatic configuration allows also for an independent increase 
of $k_\mathrm{F}$ due to the higher achievable densities. 
The resulting $T_\mathrm{BCS}/T_\mathrm{F}$  are within the 
explorable range which corresponds, as seen in the lower panel of
Fig. 3, to $T/T_\mathrm{F} \geq 2 \cdot 10^{-2}$. 
The presence of a superfluid state could be evidenced by using the 
same blue-detuned beam used to deconfine the bosons as a mechanical 
stirrer for the fermion cloud. Thus, in analogy to already performed 
experiments on Bose condensates, one could look at a finite
threshold for the onset of a highly dissipative regime \cite{Raman} 
or of a drag force \cite{Onofrio1}.  
The stirring of the Fermi gas occurs in the presence of both a bosonic
thermal cloud and a Bose condensed component. These last give rise to
heating at all stirring velocities \cite{Raman1} and at a critical
velocity lower than for the one expected for the Fermi superfluid, respectively. 
However, due to their low density in the latest cooling stage, the
contributions to the heating induced by the stirring of
the Bose components are much smaller than the Rayleigh heating.
To discriminate against the bosonic cloud background one could take 
advantage of the recently proposed manipulations of an ultracold 
cloud with Raman beams by creating a directional 
critical velocity for the superfluid Fermi component \cite{Higbie}.
An alternative signature for superfluidity consists in looking at
the bulge in the density profile predicted below 
$T_\mathrm{BCS}$ \cite{Chiofalo}. 
Here, again, the presence of a thermal cloud for the bosons makes 
this background simpler to discriminate against any fermion
superfluidity signature due to the well controllable 
Gaussian-shaped profile of the former, its weaker interactions 
with the Fermi gas \cite{Amoruso}, and the broader Thomas-Fermi
profile of the Bose-condensed component caused by the shallower 
confinement.

In conclusion, our analysis of evaporative and sympathetic cooling 
in a two-color optical dipole trap shows that a deep Fermi degeneracy 
regime can be achieved by efficiently exploiting the cooling
capability of a Bose gas with large heat capacity. 
The fact that the cooler need not  be in the condensed phase gives
also larger flexibility for choosing the Bose species. 
One could reconsider the use of $^{133}$Cs which, due to its large
mass and small recoil temperature, can be efficiently cooled to very 
low temperatures by purely optical means, therefore ensuring robust 
conditions in terms of temperature and heat capacity to start 
evaporative cooling \cite{Mudrich}.  
More generally, the presence of a larger thermal component for the Bose gas 
does not interfere significantly with the Fermi component, rather 
it allows for a more accurate thermometry and a more controllable 
background against possible signatures of the fermion superfluid phase.

\begin{acknowledgments}
We thank S. Gupta, Z. Hadzibabic, and L. Viola for
useful discussions.  This work was supported in part by
Cofinanziamento MIUR protocollo MM02263577\_001. 
\end{acknowledgments}


\begin{references}
\bibitem{NOTE} We assume that the presence of the Fermi gas does not
affect significantly evaporative cooling of the Bose gas. 
This requires that the potential energy depth of fermions 
$\Delta U_\mathrm{f}$ is larger than that of 
bosons $\Delta U_\mathrm{b}$, and that $C_\mathrm{b}/C_\mathrm{f} \gg 1$.
\end{references}
\end{document}